\documentclass[12pt]{iopart}

\usepackage{graphicx}
\usepackage{epsfig}
\usepackage{graphics}
\usepackage{bm}
\usepackage{psfrag}
\usepackage[T1]{fontenc}

\newcommand{\da}{\downarrow}
\newcommand{\ua}{\uparrow}
\usepackage{color}

\newcommand{\bessel}{\mathcal{J}}
\newcommand{\ket}[1]{\left | #1 \right\rangle}
\newcommand{\bra}[1]{\left \langle #1 \right |}
\newcommand{\eqref}[1]{$\left ( \ref{#1} \right)$}

\begin{document}
\title{Rice-Mele model with topological solitons in an optical lattice}
\author{Anna Przysi\k{e}\.zna$^{1,2,3}$ Omjyoti Dutta$^{3}$, and Jakub Zakrzewski$^{3,4}$}
{ 
\address{$^1$ Institute of  Theoretical Physics and Astrophysics, University of Gda\'nsk, Wita Stwosza 57, 80-952 Gda\'nsk, Poland }
\address{$^2$ National Quantum Information Centre of Gda\'nsk, Andersa 27, 81-824 Sopot, Poland}
\address{$^3$ Instytut Fizyki imienia Mariana Smoluchowskiego, Uniwersytet Jagiello{\'n}ski, ulica \L{}ojasiewicza 11, PL-30-059 Krak\'ow, Poland }
\address{$^4$ Mark Kac Complex Systems Research Center,
Uniwersytet Jagiello\'nski, Krak\'ow, Poland }
}

\date{\today}

\begin{abstract}
Attractive ultracold fermions trapped in a one-dimensional periodically shaken optical lattice are considered.
For an appropriate resonant shaking, dimerized structure emerges for which the system realizes paradigmatic  physics described by  Rice-Mele model. Emergent nature of the system together with  density fluctuations or controlled modifications of lattice filling allow for creation of defects. 
Those defects  
lead to topologically protected localized modes carrying the fractional particle number. 
Their possible experimental signatures are discussed.
\end{abstract}

\maketitle

\section{Introduction}
Conducting polymers \cite{Su} are particularly interesting one-dimensional systems due to their unusual topological properties characterized by a non-trivial Zak phase \cite{Zak,Niu}, degenerate ground states, topological solitons \cite{SSH} and a fractional charge \cite{Jack}. 
Such polymers can be described in a simple manner by fermions moving in a lattice with dimerized tunneling amplitudes \cite{SSH, RM}. The corresponding models can be constructed with ultracold atoms in optical lattices which give unprecedented tunability and control over the system parameters.
Recently, ultracold bosons in optical superlattices were used to  prepare experimentally \cite{RMBloch}  a model of conducting polymers (namely Rice-Mele (RM) model \cite{RM}, one of the  simplest 1D models of nontrivial topology) and the corresponding Zak phase was measured. In a parallel work
topological edge states in  a similar potential were studied theoretically \cite{ssh_superlattice}.

Models based on superlattice potentials are relatively  easy to realize in experiment, however, they have also some limitations. An optical lattice potential is typically defect-free due to its origin. That creates a difficulty in realizing topological solitons. 
Such solitons typically emerge on defects that are the domain walls between topologicaly distinct phases.   
Forming the signature of nontrivial topology, they are the essence of RM model.

In the present paper, we show how to realize RM model with controlled defects using a system of  attractive ultracold fermions \cite{Chin, Strohmaier, Hacker} in a simple shaken one-dimensional optical lattice. 
Shaking, i.e.  periodic driving of system parameters (e.g. the optical potential depth or position), has been successfully implemented in cold atomic systems  in order to induce various effects  \cite {shake0,shaking1, shaking2,shaking3,shaking4,shaking5} following the seminal proposition \cite{Andre}.
In \cite{triangle} we have shown that such a shaking  combined with attractive interactions in two-dimensional triangular lattice can result in an emergent  Dice  structure with topological properties. 
Here, we show that in the case of a one-dimensional system there exist a regime of parameters where atoms self-organize into   a dimerized  structure. The  ground state is then two-fold degenerate. The corresponding states represent  two topologically distinct dimerized configurations.  Due to the emergent nature of the dimerized state, both configurations, separated by domain walls, may be simultaneously present in the lattice. Moreover, by controlling the filling fraction, impurities may be added to the configurations. 
Such defects -- domain walls and impurities --  naturally  give rise to topologically protected solitons or bound states with a fractionalized particle number.


\section{ System }
 Our system consists of two-species (denoted as $\downarrow$, $\uparrow$) fermionic  mixture trapped in an optical lattice potential 
$V_{\mathrm{latt}}= V_{\parallel}\sin^2(\pi x/a) + V_{\perp}(\sin^2(\pi y/a) + \sin^2(\pi z/a)) $, where $a$ is the lattice constant. For $V_{\perp}\gg V_{\parallel}$  the system is effectively one dimensional. To control the system we use
a familiar lateral (horizontal) lattice shaking \cite{Andre}. Importantly, however, we  introduce also periodic changes in the potential depth which we call here {\it vertical shaking}: $V_\parallel=V_0+\delta V_0 \cos \omega t$. $\delta V_0$  is an amplitude of the lattice depth shaking and $\omega$ -- the frequency, common to the lateral and vertical shaking.  
We assume fermionic species of equal mass, $M$, with  different fillings: $n_{\downarrow}\approx 1$ and $n_{\uparrow}\approx 1/2$. The interaction between atoms of different species is assumed to be attractive.
In effect fermions of different spin tend to pair creating composites \cite{Dut}  with the density given by the minority $\ua$-fermion density  $n^{\ua}$. We include $p$-bands in the model and effectively have the composites that occupy $s$-bands and excess $\downarrow$-fermions that may occupy both $p$ and $s$-bands.

The Rice-Mele model \cite{RM} contains two essential ingredients: two types of sites and asymmetric couplings between them. The former is realized in our model by a density-wave self-arranged configuration of composites. Such a configuration is energetically favorable when intra-band tunnelings are switched off by appropriate adjustment of shaking amplitude while making density-dependent inter-band tunneling resonant by adjusting the shaking frequency. To obtain asymmetric coupling with nearest neighbours the additional vertical shaking is necessary with an appropriate phase shift with respect to standard lateral shaking.
This phase difference breaks left--right symmetry of the problem.

To write the effective Hamiltonian of the model, we construct the time dependent Hamiltonian, $H(t)$ and  average it in time \cite{Andre}. The minimal Hamiltonian of our system contains tunnelings, density induced tunnelings, renormalized interactions and shaking: 
 $\hat{H}=\hat{H}_{\rm tun}+\hat{H}_{\rm dit}+\hat{H}_{\rm int}+\hat{H}_{\rm sh}(t),$ where: 
\begin{eqnarray}
\label{Ht}
\hat{H}_{\rm tun}&=& J_0\sum_{ \langle {ij}\rangle}  \left [\hat{s}^\dagger_{i}\hat{s}_{j}+ \hat{s_\ua}^\dagger_{i}\hat{s_\ua}_{j}\right]+ J_1 \sum_{ \langle {ij}\rangle} \hat{p}^\dagger_{i}\hat{p}_{j},\nonumber\\
\hat{H}_{\rm dit}&=&
\sum_{\langle {ij}\rangle}\left [ T_0 \hat{s_\ua}^\dagger_i(\hat{n}^{}_ {i}
+\hat{n}^{}_ {j})\hat{s_\ua}_{j}+T_1{\hat p_{i}}^{\dagger}(\hat{n}^\ua_ {i}
+\hat{n}^\ua_ {j})     \hat{p}_{j}
 \right. \nonumber\\
&\quad&+\left. 
 T_{01}((j-i){\hat p_{{i}}}^{\dagger}\hat{n}^\ua_ {i} \hat s_{{j}}^{} + h.c )  \right ],\\
\hat{H}_{\rm int}&=& U_{0}^{}\sum_{i}\hat{n}^\ua_i \hat{n}_i+ U_{1}^{}\sum_{i}\hat{p}^\dagger_i\hat{p}_i\hat{n}^\ua_i+E_1\sum_{i}\hat{p}^\dagger_i\hat{p}_i, \nonumber \\
\hat{H}_{\rm sh}(t)&=& K \cos \omega t \sum_{j}  j  (  
\hat{n}^\ua_j + \hat{s}^\dagger_j\hat{s}_j+
\hat{p}^{\dagger}_j\hat{p}_j ) 
+  \delta E_1\cos (\omega t +\varphi) \sum_{i}{\hat p_{ {i}}}^{\dagger}\hat{p}_{{ {i}}}^{}.\nonumber
\end{eqnarray}
Here, 
$\hat{s}^{\dagger}_{{i}}, \hat{s}_{{i}}^{}$, $\hat{p}^{\dagger}_{ {i}}, \hat{p}^{}_{ {i}}$ are creation and annihilation operators of $\downarrow$-fermions in the $s$- and $p$-bands respectively, while $\hat{s}^{\dagger}_{\ua{i}}, \hat{s}_{\ua{i}}^{}$ are $s$-band creation and annihilation operators for $\ua$-fermion.  Accordingly, $\hat n_i, \hat n^p_i,$ and $\hat n^\ua_i$ are the corresponding
number operators.
In the on-site interaction  Hamiltonian, $\hat H_{\rm int}$, the renormalized self-energy of the composites is denoted by $U_0$, the on-site renormalized interaction between the composite and an excess $\downarrow$-fermion is characterized by $U_{1}$, and $E_1$ is the energy of the $p$-band. {Modulus of negative $U_0$ values is the largest interaction energy scale assumed in the model with $|U_1|<|U_0|$ (this follows from properties of Wannier functions, see Appendix A) . Attractive interactions between two species lead then to creation of composites in the $s$ band.  All the tunneling amplitudes are assumed to be much smaller than $\omega$. The range of $\omega$  values will be set by the required resonance condition (\ref{neweqnumber})}

$\hat H_{\rm tun}$ corresponds to standard tunnelings with  amplitudes  $J_0$ and $J_1$ for the $s$ and $p$  bands while  $\hat H_{\rm dit}$ describes
 often neglected density induced tunnelings (\cite{Dutta11,Mering,dirk12,ropp} with amplitudes $T_i$.
 Observe that $\uparrow$-tunneling from site  $i$ to site $j$ may happen only when  there is a composite
 on site $i$ and   a free $\downarrow$-fermion in the $s$-band on site $j$. In effect, this tunneling creates a composite on site $j$. 
In the case of $\downarrow$-fermions analogical situation does not take place because there are no free $\uparrow$-fermions. 
The last -- most important for {the mechanism discussed later} -- term of $H_{dit}$ couples $s$ and $p$ levels and describes process occurring  when composite-empty site adjoins composite-occupied one. (for details see  Appendix A).

$\hat H_{\rm sh}(t)$ is a time-periodic Hamiltonian with $K$ denoting the amplitude of the lateral lattice shaking while
$\delta E_1$ denotes the strength of time-variation of single-particle energy in the $p$-band which is induced by periodic driving of the lattice depth  while $\varphi$ is a relative phase between the lateral and vertical drivings. Additional effects due to the vertical shaking that are negligible for moderate $\delta V_0$ are discussed in Appendices.

Next, we describe the averaging process (see Appendix B for more details). 
First, we apply the unitary transformation, $\hat{U}=\exp [- {\rm i} {\hat H}_{\rm int} t - {\rm i} \int^t_0 {\hat H}_{\rm sh}(t')dt']$ that gives us a new Hamiltonian 
$\hat H'=\hat{U}^\dagger \hat H \hat{U} - {\rm i}\hat{U}^\dagger [d_t \hat{U}]$.  In comparison with $\hat H$ the shaking and the on-site interaction parts are removed by $\hat U$ while the tunneling part is dressed in  oscillating terms. We aim at the situation when inter-band density dependent $sp$ tunneling makes the dominant tunneling contribution. Accordingly,
 we assume the resonant condition 
 \begin{equation}
 \label{neweqnumber}
 E_1+U_1=N\omega + 2\Delta,
  \end{equation}
 where $N$ is integer and $\Delta \ll \omega$ is the detuning and we time average the Hamiltonian. In the process fast oscillating   terms $\sim 1/\omega$
are neglected.  Necessarily the shaking frequency $\omega$ is chosen  large compared to all the tunneling amplitudes. 
We obtain  {the effective Hamiltonian} $H_{\rm eff}=\tilde {H}_{\rm tun} +\tilde {H}_{\rm dit}+\tilde {H}_{\rm ons}$,
with
\begin{eqnarray}
\label{Heff}
\tilde {H}_{\rm tun} &=& \tilde J_0\sum_{ \langle {ij}\rangle}  \left [\hat{s}^\dagger_{i}\hat{s}_{j}+ \hat{s_\ua}^\dagger_{i}\hat{s_\ua}_{j}\right]+ \tilde J_1 \sum_{ \langle {ij}\rangle} \hat{p}^\dagger_{i}\hat{p}_{j}\nonumber\\
\tilde {H}_{\rm dit} &=&
\sum_{\langle {ij}\rangle}\left [ \tilde T_0 \hat{s_\ua}^\dagger_i(\hat{n}^{}_ {i}
+\hat{n}^{}_ {j})\hat{s_\ua}_{j}+\tilde T_1{\hat p_{i}}^{\dagger}(\hat{n}^\ua_ {i}
+\hat{n}^\ua_ {j})     \hat{p}_{j}  \right]\\
&+& \tilde{T}^-_{01} \sum_{i} \left(\hat s_{2i}^{\dagger}\hat p_{2i-1}^{}+h.c.\right)
 - \tilde{T}^+_{01} \sum_{i} {\left(   \hat s_{2i}^{\dagger}\hat p_{2i+1}^{}  +h.c.\right)}\nonumber\\
\tilde {H}_{\rm ons}&=& \Delta  \sum_{\langle {ij}\rangle}{\hat p_{i}}^{\dagger} p_{i}^{}- \Delta  \sum_{\langle {ij}\rangle}{\hat s_{ {i}}}^{\dagger}s_{ i}.\nonumber
\end{eqnarray}
 The intra-band tunneling parts above are modified in the standard manner \cite{Andre}:  $\tilde J_{l} =\mathcal{J}_0\left(\frac{K}{\omega}\right)J_{l}$ (as well as  $\tilde T_{l} =\mathcal{J}_0\left(\frac{K}{\omega}\right)T_{l}$) for $l\in \{0,1\}$, where
$\mathcal{J}_0\left(\frac{K}{\omega}\right)$ is the ordinary Bessel function of order zero.
In the case of the inter-band part, time averaging brings us, however, a new  effect. The inter-band hopping is modified by the Bessel function of order $N$ with different amplitudes  depending on the {\it direction} of this process ($+$ or $-$): $\tilde T_{01}^\pm =\mathcal{J}_N\left( A^\pm/\omega\right)T_{01}$ where $A^\pm =\sqrt{(K\pm\delta E_1 \cos\varphi)^2+K^2 \sin^2\varphi}$. These amplitudes depend on the relative phase of the drivings, $\varphi$, which, we believe, can be controlled in real experiments with a good precision. 
The  detuning, $\Delta$,  leads to residual  on-site potential $\tilde H_{\rm ons}$. 

Now  we can tune the hopping parameters. The intra-band amplitudes may be made very small by choosing $K/\omega$ such that $\bessel_0(K/\omega)\approx 0 $. For slightly different $K/\omega$ $ss$ hopping remains negligible (so the composites may be still considered as immobile) while  the typically much larger  $pp$ hopping start to plays a role and has to be taken into account. At   the same time the inter-band hopping is large since it depends on Bessel functions of order $N\ne 0$ .

 From now on we set the recoil energy, $E_R=h^2/(8Ma^2)$, as an energy unit and consider the ground state structure of $H_{\rm eff}$  on an exemplary case of lattice depths $V_0=8$, $V_\perp=25$ and interaction strength $\alpha=a_s/a=-0.1$ (with $a_s$ being the (negative) scattering length). We choose the vertical shaking to be in phase with the lateral one  ($\varphi=0$),  giving $\mathbf {\tilde T^-_{01}>\tilde T^+_{01}}$.
In the region with dominant inter-orbital tunneling, we expect that the ground state is given by the density wave configuration (DW) with every second side occupied by composites. {The reason for that is quite simple.  In such a configuration neighboring sites  may contain excess fermions only and the $sp$ tunneling lowers the energy for such a situation. On the other hand if two composites reside in the consecutive sites then the inter-band $sp$ tunneling would break one of the composites (which costs the energy) while intraband $pp$ tunneling for excess fermions is assumed small so it cannot lower the energy.}
   To confirm that prediction, we
first assume the composites to be immobilized due to the negligible value of { $\mathbf {\tilde J_{0}}$ and $\mathbf {\tilde T_{0}}$ when compared to other hopping amplitudes } in the whole regime under consideration.  
Then finding the ground state configuration  boils down to solving the single particle Hamiltonian for a group of all possible configurations of $\hat{n}^\ua_i$ (that determines composites distribution). Within the approximation of immobilized $\ua$-fermions, we can replace their number operators by c-numbers ${n}^\ua_i=1,0$  depending on the presence or absence of the fermion on site $i$.
Since the search space grows exponentially with the number of sites, it quickly becomes too large for the exact diagonalization. Thus  we apply Simulated Annealing
\cite{SA} to  find the dependence of lowest energy configuration on  $K/\omega$.  Our calculations have been performed for the lattice of 24, 40 and 60 sites and the obtained results do not depend on the number of sites.

\begin{figure}
\begin{center}
\hspace{-0.5mm}
\includegraphics[width=120mm]{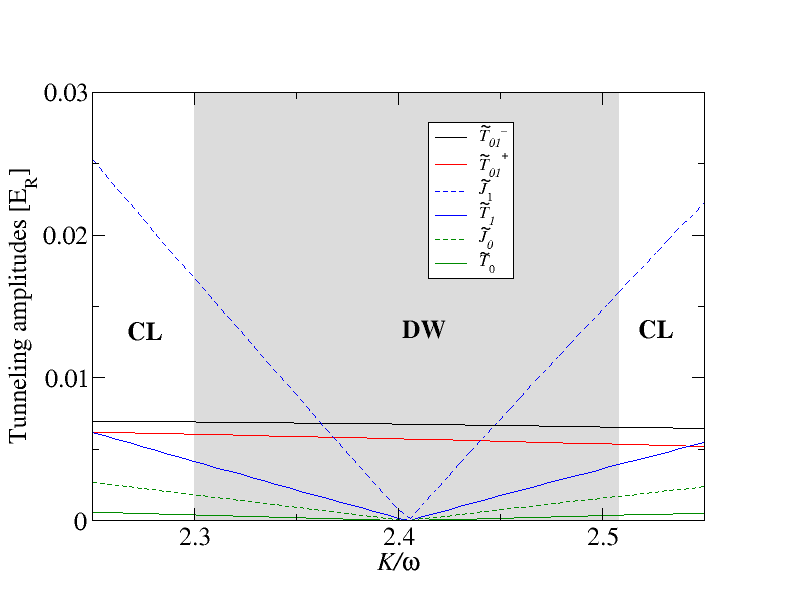}
\caption{\label{JJJ} Dependence of different hopping amplitudes on shaking parameter $K/\omega$ for the exemplary system of $V_0=8$, $V_\perp=25$, $\alpha=-0.1$, $N=1$, $\varphi =0$. Energies are expressed in recoil energy units as discussed in the text. Gray area marks  the interval of $K/\omega$ in which we obtain density wave (DW) structure of $\ua$-fermions- outside this region { composites form a cluster (CL).}}
\end{center}
\end{figure}

The configurations obtained for $\Delta = 0$ are shown in Fig. \ref{JJJ}. 
We find indeed two possible configurations of composites 1) { clustered phase (CL) where the composites cluster together  with the rest of the lattice being empty} or 2) DW phase (the shadowed region) where we have alternation of occupied and empty sites. We see that DW structures occur for the shaking parameter, $K/\omega$, for which $|\tilde T_{01}^-|+|\tilde T_{01}^+|  \geq  |J_{1}|\bessel_0(K/\omega)$. Simulations performed for other values of $\Delta$ gave similar dependence of the ground state configuration on $K/\Omega$ with the  region  of DW configuration broadening slightly for $\Delta >0$. The obtained phase diagram is stable under small fluctuations in the fermion densities for both $\ua$- and $\da$-fermions -- the DW structure is preserved with some sparse defects appearing. 

\section{Emergent Rice-Mele model }
 When the DW-configuration of the composites minimizes the energy of the system, the dominant hopping process is the inter-band $sp$-one and the effective Hamiltonian for the excess $\downarrow$-fermions corresponds to the Rice-Mele model \cite{RM},
\begin{eqnarray}
H_{\rm DW} &=&\tilde{T}^-_{01} \sum_{i} \left(\hat s_{2i}^{\dagger}\hat p_{2i-1}^{}+h.c.\right)
 - \tilde{T}^+_{01} \sum_{i} {\left(   \hat s_{2i}^{\dagger}\hat p_{2i+1}^{}  +h.c.\right)}\nonumber\\
 &+&\Delta  \sum_{\langle {i}\rangle}{\hat p_{2i+1}}^{\dagger} p_{2i+1}^{}- \Delta  \sum_{\langle {i}\rangle}{\hat s_{ {2i}}}^{\dagger}s_{ 2i}.
\label{Hdw}
\end{eqnarray}
From now we shall drop the $\tilde{ \ }$ sign over tunneling amplitudes as we shall consider effective tunnelings only restricting to (\ref{Hdw}). The above Hamiltonian describes a perfect lattice without defects. However, if the defects in the lattice are sparse with comparison to the edge mode length (see section \ref{edge}) and  we tune the shaking to make intra-band hopping  small, then each of the domains may be separately described by the Rice-Mele Hamiltonian \eqref{Hdw}.  Otherwise, a proper description of the system requires  including { also} intra-band tunnelings.

\begin{figure}
\begin{center}
\includegraphics[width=120mm]{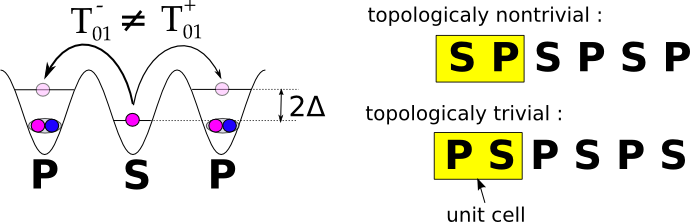}
\caption{\label{system}  (a) Pictorial representation of the system described by $H_{DW}$. $\ua$-fermions are denoted by blue circles while  $\da$-fermions are denoted by pink ones. Arrows indicate direction of tunneling. Sites occupied by the composites are denoted by {\bf P}, sites not occupied by the composites are denoted by {\bf S}.  Two possibilities of the DW configurations (b): ``SP'' at the top and ``PS'' at the bottom. Unit cells are marked with yellow rectangles.  }
\end{center}
\end{figure}

To write the Hamiltonian in the momentum space we specify a unit cell to contain two neighboring sites of which only one is always occupied by a composite. 
Such a unit cell can be chosen in two different ways depending whether the composite { resides in the first (we call such a configuration ``PS'')  or the second site (``SP'')} of the open chain (see Fig.~\ref{system}).  
As expected for RM model \cite{RM}  these two choices of the unit cell give rise to topologically distinct states. When written in the momentum space, the Hamiltonian \eqref{Hdw} reads:
\begin{equation}
H_{DW}^\pm(k)=\pm (T_{01}^\mp-T_{01} ^\pm \cos(2ka)) \sigma_x+
T_{01}^\pm \sin(2ka) \sigma_y
-\Delta \sigma_z,
\label{Hpm}
\end{equation}
where the $H_{DW}^+(k)$ corresponds to "PS" configuration,  $H_{DW}^-(k)$ corresponds to the "SP"-one,  and $\sigma_{x,y,x}$ are Pauli matrices. 
The dispersion relations are the same for both configurations:
\begin{equation}
\epsilon_{\pm}(k)=\pm\sqrt{\Delta^2+(T_{01}^{+}+T_{01}^{-})^2+4T_{01}^{+}T_{01}^{-}\cos^2(ka)}.
\end{equation}

Topologically distinct configurations are characterized by different Zak phases \cite{Zak} (i.e. Berry phases acquired across the Brillouin zone). Zak phases of particular states depend on the choice of the unit cell, their difference forms an invariant of the system. 
The Zak phase is given by \cite{Zak,Niu}:
\begin{equation}
\phi_{Zak}=i\int _{-\pi/2} ^{\pi/2} \bra{u_k}\partial_k \ket{u_k} dk,
\end{equation}
where $\ket{u_k}$ are Bloch functions of the system i.e. eigenfunctions of the Hamiltonian \eqref{Hpm}. For $\Delta=0$, when the Hamiltonian is equivalent to the SSH model \cite{Su}, we obtain $\phi_{Zak}^{SP}-\phi_{Zak}^{PS}=\pi$. This indicates that "SP" and "PS" phases are topologically distinct -- one of them must be nontrivial. For nonzero $\Delta$ we obtain fractional (in units of $\pi$) Zak phase differences changing from $\pi$ to $1.67\pi$ for $\Delta\in[0, 0.002\omega]$. To determine which configuration has a nontrivial topology, we investigate the existence of edge modes.

\section{Localized modes \label{edge}}

 In our model, defects arise naturally due to the emergent nature of the DW structure. As discussed in \cite{Dut}, the time scale required to reach a particular DW lattice configuration is set by the minority component tunneling rate. Subsequently the timescale to form the entire DW configuration is governed by the corresponding Lieb-Robinson bound \cite{Lieb72}.
When the time of creation is not sufficiently long, smaller regions of different DW configurations, separated by domain walls, will be created. 
Moreover, due to number fluctuations present for trapped atoms, the composites will not be exactly at half-filling. 
Any deviation from this filling will result in a defect in the form of a vacancy or a filled site. 

\begin{figure}
\begin{center}
\includegraphics[width=120mm]{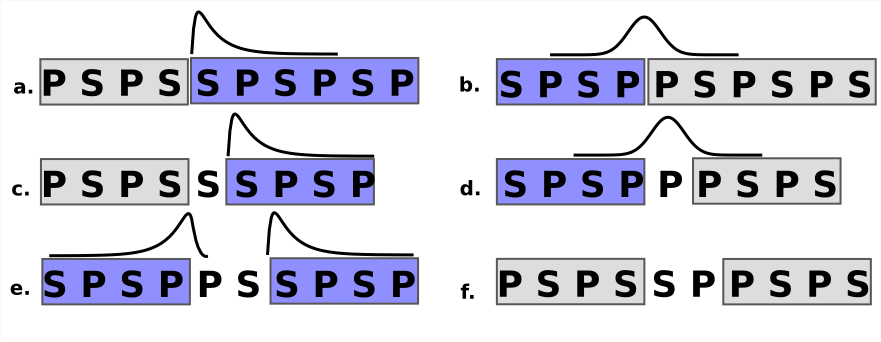}
\caption{\label{defects}  Possible defects in the system and localized modes that may grow on them. With light gray (blue) background we mark topologically (non)trivial phase. On domain walls (panels a,b,c,d)  we have one localized  state:
 vanishing sharply on the edge and localized on the side of nontrivial phase when there is no tunneling possible between sites separating domains (panels a,c);  vanishing exponentially  on both sides of the domain wall if there is hopping possible through the wall (panels b,d). If there is a small impurity within one phase we can either have  two modes on both sides of the impurity if the phase is nontrivial (panel e), or no modes for a trivial phase (panel f)}.
\end{center}
\end{figure}

Both kinds of impurities -- domain walls and lattice defects -- give rise to topologically protected localized modes \cite{Shen} (visualized in  Fig.~\ref{defects}). If we tune the shaking to make both $ss$ and $pp$ tunnelings negligibly small, then on domain walls we effectively create open boundary conditions. 
 This will result in an appearance of edge modes in SP configuration when  $T^-_{01}>T^+_{01}$ and in PS configuration otherwise. These modes vanish sharply on the edges (compare Fig.~\ref{defects}a and Fig.~\ref{defects}c ) indicating that the configuration has a nontrivial topology. Let us here focus on the case when $T^-_{01}>T^+_{01}$.
The edge modes have energies $\pm\Delta$ and {in the continuum limit, their eigenvectors are given by the spinor: $(\psi_s(x), \psi_p(x))$}
Depending on which edge we are, setting $x=0$ on the boundary, we get:  the edge mode on the left end (the one ending with S site) with energy $-\Delta$ where $\psi_s(x)=A(e^{-\lambda^+ x}-e^{-\lambda^- x})$, $\psi_p(x)=0$  and the edge mode on the right end (the one ending with P site) with energy $\Delta$, where $\psi_s(x)=0$, $\psi_p(x)=A(e^{\lambda^+ x}-e^{\lambda^- x})$ with
\begin{equation}
\lambda^\pm=\frac{T_{01}^-\pm \sqrt{T_{01}^-(2T_{01}^+-T_{01}^-)}}{2T_{01}^-},
\end{equation}
and $A$ being the normalization constant. {A detailed and tutorial discussion of edge modes in the dimer model is given in \cite{Ullmo}. } 

When tuning $K/\omega$ further from the zero point of the Bessel function, the $ss$ hopping is still negligible, but the $pp$ hopping becomes significant. Therefore, on those boundaries that are separated by P-sites (Fig.~\ref{defects}b and Fig.~\ref{defects}d), particles can tunnel through the boundary and the mode vanishes exponentially on the both sites giving topological solitons with energy $\Delta.$ 
Defects occurring inside the "SP" configuration (Fig. \ref{defects}e) give rise to two localized modes  on both sides of the impurity. 
Depending on $ss$ and $pp$ tunneling rates they may end sharply on the boundary or smoothly vanish inside the defect. The width of the edge states depends on the hopping amplitudes and can be changed by tuning the value of $K/\omega$. For $K/\omega=2.3$ the edge state is about 15 lattice sites long and it becomes narrower with higher values of $K/\omega$. 

 Defects present in the lattice are associated with local changes of fermion number by fraction, $N_{\rm frac}=f_0/\pi$ where $f_0=\tan^{-1} \left[|T_{01}^{+}-T_{01}^{-}|/2\Delta \right]$ \cite{RM} at zero temperature. At finite temperatures, $T\ll \Delta$,  the corresponding fractional fermion number for the localized mode is given by the thermal expectation value \cite{frac,frac1}, $N^T_{\rm frac}= f_0/\pi-sgn(f_0) \exp[-\Delta/T]$, where $sgn()$ is the sign function. For a typical value of $\Delta=0.01E_R$ the relevant temperature  is in the nanokelvin regime for  ${}^{40}K$.

Finally let us briefly comment on the dynamics of solitons. The solitonic localized modes are pinned to the defects. Their dynamics is affected by tunnelings as well as thermal excitations at finite temperature.   A single density dependent tunneling event [given by $T_0$ term in the original Hamiltonian (\ref{Ht})] will change the "PS" pair to "SP". This corresponds to a motion of the defect and thus the motion of the soliton localized on the edge(e.g. compare Fig.~\ref{defects}a or Fig.~\ref{defects}b). While by adjusting the frequency we minimize the influence of such processes they will be still partially present due to, e.g., a frequency mismatch with respect to the exact zero of the appropriate $\mathcal{J}_0 (K/\omega)$ Bessel function or higher order terms,  discussed in the next Section. One may envision also that once the system is formed, a frequency/amplitude of shaking is changed a little to stimulate the motion of defects.

 Consider, however, the situation represented in Fig.~\ref{defects}e. The single "PS" to "SP" tunneling will remove two defects - that would correspond to collisional anihilation of two localized modes. The stable ``solitonic'' solution correspond thus to situations with
 well separated defects that  cannot easily get removed by tunnelings.

\section{Experimental realization and probing \label{exp}}
Let us first discuss time scales needed to realize the described system experimentally. The time for the formation of the crystal  is determined by tunneling rate of $\ua$-fermions and bounded from above by loss rates. On an example of ${}^{40}K$ we estimate the tunneling time to be of the order of 10ms (compare Fig.~1) except at the close vicinity of $K/\omega=2.4$. Close to $K/\omega=2.4$ when the tunneling in the averaged Hamiltonian vanishes there will be still residual higher-order tunneling.  Its non-resonant effect can be estimated in analogous way to Bloch-Siegert shifts in quantum optics
\cite{AllenEberly} being proportional to $J_0^2/4\omega$ which again for  ${}^{40}K$ is about 10ms.
The formation of the crystal will take a few tunneling times. The precise estimate would require full dynamical calculation of the crystal formation which is beyond the scope of the present paper.

The shorter time will lead to numerous defects. Number of defects is dependent also on the temperature. For the nanokelvin regime and assuming a sufficient preparation time, the  number of defects (which can be calculated \cite{Mermin} comparing the energy of creating defect with temperature assuming canonical ensemble) will be of the  order of 1\% -- that allows one to obtain well separated edge states. 
Having the system prepared, we may probe its topological properties. The Zak phase can be measured experimentally in a way it has been proposed in Ref. \cite{RMBloch} -- with application of coherent Bloch oscillations combined with Ramsey interferometry. 
{At exact half filling of excess fermions the topologically non-trivial band will be filled and the standard time of flight method cannot detect the Zak phase\footnote{We are grateful to one of the referees for pointing that out}. Slightly lower filling of excess fermions does not affect the DW structure of composites (so Rice-Mele model is applicable). Then such a measurement of Zak phase should be possible. 
For this measurement also the number of defects should not be too large.} However, to detect the localized states, defects can be helpful. Localized states can be observed with photo-emission spectroscopy  \cite{Jin1} with time-of-flight where number of edge modes in our system results in an increased peak intensity near  zero momentum making the signal less susceptible to noise.
At half filling, the only localized states that can be occupied are those of the negative energy. 
Fermion number fractionalization can be probed on defects with application of the single site imaging. 

\section{Conclusions} 
We have shown that a combination of shaking and attractive interactions in 1D optical lattice can give rise to a topologically nontrivial system. We have used standard lateral shaking but also introduced an additional vertical shaking. Together, they result in a dimerized tunneling structure. Moreover, by tuning the onsite energy slightly out of the resonance we can induce the staggered potential. By controlling the filling, we have shown further the presence of topologically protected localized modes. We believe that such modes can be experimentally verified at accessible temperatures.  

\section*{Acknowledgments} 

Numerical calculations were performed at the Academic Computer Center in Gda\'nsk.
This work was realized under National Science Center (Poland) project No. DEC-2012/04/A/ST2/00088. A.P. is also supported by the International PhD Project "Physics of future quantum-based information technologies", grant  MPD/2009-3/4 from Foundation for Polish Science and by the University of Gdansk grant BW 538-5400-B524-14.

\section*{Appendix A}\label{appa}

We derive the minimal model in a standard manner starting from  many body Hamiltonian of dilute gas of atoms in a second quantization representation \cite{Lew,ropp}. We consider two species (denote by $\ua$-fermions and $\da$-fermions) of
equal masses which can occupy the lowest band. The $\da$-fermions have occupation close to unity, for them we consider also the excited, $p$ orbital. Different species undergo contact interactions. The parameters in the Hamiltonian (1)
in the main text  are given by integrals of Wannier functions  $\mathcal{W}^{0(1)}_{\mathbf{i}}(x,y)$ on $s$($p$)-bands, where $i$ is a site index. 

Specifically the single particle $ss$ and $pp$ hoppings do not depend on the type of species and read
\begin{eqnarray}
J_{0} &=& \int \left[\mathcal{W}^{0}_{i}(x)\right]^* H_{\rm latt}\mathcal{W}^{0}_{i+1}(x)dx,\nonumber\\
J_{1} &=& \int \left[\mathcal{W}^{1}_{i}(x)\right]^* H_{\rm latt}\mathcal{W}^{1}_{i+1}(x)dx,
\end{eqnarray}
where $H_{latt}=-\frac{\partial^2}{\partial x^2}+V_0\sin^2(\pi x/a)$ is a single particle Hamiltonian for a static lattice.
Observe the lack of two in the kinetic energy as we work in recoil units. The contact interactions between different species lead
to density induced tunnelings \cite{Dutta11,Mering,dirk12,ropp}.  The corresponding part of the Hamiltonian may be expressed as
\begin{eqnarray}
\label{Htsup}
\hat{H}_{\rm dit} &=&
\sum_{\langle {ij}\rangle}\left [ T_{01}((j-i){\hat p_{{i}}}^{\dagger}\hat{n}^\ua_ {i} \hat s_{{j}}^{} + h.c )
+T_1{\hat p_{i}}^{\dagger}(\hat{n}^\ua_ {i}
+\hat{n}^\ua_ {j})     \hat{p}_{j} \right.\nonumber\\
&+&T_1' \hat{s_\ua}^\dagger_i(\hat{n}^{p}_ {i}
+\hat{n}^{p}_ {j})\hat{s_\ua}_{j}
+\left. T_0 \hat{s_\ua}^\dagger_i(\hat{n}^{}_ {i}
+\hat{n}^{}_ {j})\hat{s_\ua}_{j} 
+ T_0' \hat{s}^\dagger_i(\hat{n}^{\ua}_ {i}
+\hat{n}^{\ua}_ {j})\hat{s}_{j}   \right ],
\end{eqnarray}
where, let us recall, $\hat{s}^{\dagger}_{{i}}, \hat{s}_{{i}}^{}$, $\hat{p}^{\dagger}_{ {i}}, \hat{p}^{}_{ {i}}$ are the creation and annihilation operators of the $\downarrow$-fermions in the $s$- and $p$-bands respectively, while $\hat{s}^{\dagger}_{\ua{i}}, \hat{s}_{\ua{i}}^{}$ are $s$-band creation and annihilation operators for $\ua$-fermion.  $\hat n_i, \hat n^p_i,$ and $\hat n^\ua_i$ denote the corresponding number operators. 
Throughout the paper we assume that minority $\ua$-fermions appear in pairs only due to strong attractive interactions. 
Thus some of the processes included above vanish. In particular, the term proportional to $T_0'$ should be excluded as occupation of  $i$-site by $\ua$-fermion means that there is a $\da$-fermion occupying this site already, so Pauli principle inhibits tunneling into this site. For a different reason
$T_1'$ may also be  omitted as {  for the ground state we focus on} the occupation of $i$-site by $p$ fermion is possible energetically only if there is a composite there. 
The presence of a composite prohibits tunneling into this site of $s$-fermion. There is a possibility that a site occupied by a composite and a $p$ fermion  neigbors a site with two fermions: one in the $s$ and the other in the $p$ band. Such sites can exchange $s$-type fermion by the $T_1'$ process. One should keep in mind that $p$-fermions appear in the system only due to resonant shaking (otherwise they cost $E_1$) - their presence in neighboring sites should be a rare event.
The remaining terms form $\hat{H}_{\rm dit}$ included in the Hamiltonian (1) of the paper.
 
The amplitudes, $T$'s,  are given by integrals over four Wannier functions and take the form
\begin{eqnarray}
T_0  &=g_{1D}(\alpha) \int \left[\mathcal{W}^{0}_{i}(x)\right]^* \left|\mathcal{W}^{0}_{i}(x)\right|^2 \mathcal{W}^{0}_{i+1}(x)dx, \nonumber \\
T_1  &=g_{1D}(\alpha) \int \left[\mathcal{W}^{1}_{i}(x)\right]^* \left|\mathcal{W}^{0}_{i}(x)\right|^2 \mathcal{W}^{1}_{i+1}(x)dx, \\
T_{01}  &=g_{1D}(\alpha) \int \left[\mathcal{W}^{1}_{i}(x)\right]^* \left|\mathcal{W}^{0}_{i}(x)\right|^2 \mathcal{W}^{0}_{i+1}(x)dx, \nonumber 
\end{eqnarray}
where $g_{1D}(\alpha)$ is a renormalized 1D coupling constant \cite{Olshani} and $\alpha=a_s/a$ is the ratio of the interaction strength to the lattice spacing.
\begin{figure}
\includegraphics[width=120mm]{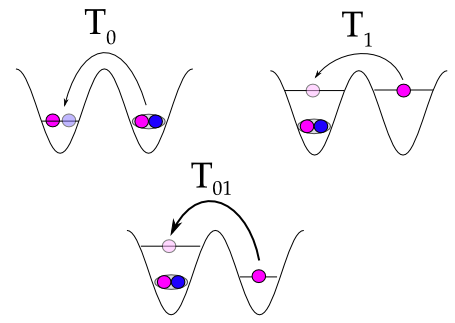}
\caption{\label{sup_fig} {Visualization of different density dependent tunneling processes present in the system. Blue and pink circles denote $\ua$-fermion  and $\da$-fermion, respectively.}}
\end{figure}

Pictorial representation of different tunneling processes is shown in Fig. \ref{sup_fig}.
Note that, since $\uparrow$ -fermions are minority fermions, they are always paired and probability of their tunneling to the $p$-band is negligibly small. That is why in case of these fermions we consider only $ss$ tunneling. 
On the other hand, the presence of $\uparrow$ -fermions (and therefore composites) stimulates $pp$ tunneling
of $\da$-fermions. 
Observe that the first term in (\ref{Htsup}) is the inter-band $sp$ hopping which has a staggered nature [reflected by $(j-i)$ sign]  and may happen when composite-empty site adjoins composite-occupied one.

Let us now discuss the on-site energies present in the Hamiltonian. The corresponding term reads:
\begin{equation}
\hat{H}_{\rm int}= U_{0}^{}\sum_{i}\hat{n}^\ua_i \hat{n}_i+ U_{1}^{}\sum_{i}\hat{p}^\dagger_i\hat{p}_i\hat{n}^\ua_i+E_1\sum_{i}\hat{p}^\dagger_i\hat{p}_i.
\end{equation}
$U_0$, $U_{1}$  are given by:
 \begin{eqnarray}
U_{0} &=&g_{1D}(\alpha) \int \left|\mathcal{W}^{0}_{i}(x)\right|^4 dx, \label{eva}\\
U_{1} &=&g_{1D}(\alpha) \int \left|\mathcal{W}^{0}_{i}(x)\right|^2 \left|\mathcal{W}^{1}_{i}(x)\right|^2dx. \label{evb}
\end{eqnarray}
$U_0$ is by far the biggest (on the modulus) energy scale and is responsible for pairing. {We assume that composites are formed in $s$ band only.  The composites could form also in the $p$ band with (negative) energy $U_0^p$ given by (\ref{eva}) with $\left|\mathcal{W}^{1}_{i}(x)\right|$ instead of $\left|\mathcal{W}^{0}_{i}(x)\right|$. Since $p$-orbitals are extended by comparison with $s$-functions, $U_0^p\approx 0.6 U_0$ for typical lattice depths (e.g. at $V_0=8E_R$ as assumed  in numerical calculations). Similarly $U_1\approx  0.4 U_0$}. Single particle energy of occupying $p$-band $E_1$ reads:
\begin{eqnarray}
E_{1} &=& \frac{1}{2} \int \left[\mathcal{W}^{1}_{i}(x)\right]^* H_{\rm latt}\mathcal{W}^{1}_{i}(x)dx
-  \frac{1}{2} \int \left[\mathcal{W}^{0}_{i}(x)\right]^* H_{\rm latt}\mathcal{W}^{0}_{i}(x)dx,
\end{eqnarray}
with the origin of the energy axis corresponding to the $s$-fermion single particle energy. {$E_1$ may be larger than $|U_0|$.}

Consider now the effects due to lateral and vertical shaking. The former is quite standard \cite{Andre} and leads a familiar  term
$$  K \cos \omega t \sum_{j}  j  (  
\hat{n}^\ua_j + \hat{s}^\dagger_j\hat{s}_j+
\hat{p}^{\dagger}_j\hat{p}_j ),
$$ where $K$ is the shaking amplitude. The vertical shaking of the lattice depth (assumed to be not too large)  causes periodic changes of single particle hoppings  $J_{z}(t)=J_{z}+\delta J_{z}\cos \omega t$ for $z=0,1$,  with amplitudes:
\begin{equation}
\delta J_{z}= \int \left[\mathcal{W}^{z}_{i}(x)\right]^* \left(\delta V_0 \sin^2\frac{\pi x}{a} \right)\mathcal{W}^{z}_{i+1}(x)dx.
\end{equation}
On  time averaging we will see that these periodic changes have negligibly small influence on the system and can be omitted. That is the reason, why they do not appear in the Hamiltonian (1) of the main text.

Next we have periodic changes in the onsite energy with amplitudes:
\begin{eqnarray}
\delta E_{1} &=& \frac{1}{2} \int \left[\mathcal{W}^{1}_{i}(x)\right]^* \left(\delta V_0 \sin^2\frac{\pi x}{a}  \right)\mathcal{W}^{1}_{i}(x)dx\nonumber \\
 &-&
  \frac{1}{2} \int \left[\mathcal{W}^{0}_{i}(x)\right]^*\left(\delta V_0 \sin^2\frac{\pi x}{a}  \right)\mathcal{W}^{0}_{i}(x)dx.
\end{eqnarray}
On the contrary to changes in the tunneling, this onsite effect is very important for the model and allow us to realize tunneling dimerization of the RM model.

\section*{Appendix B}\label{appb}

The standard time-averaging procedure  can  be obtained applying  Floquet theorem following \cite{Floquet} and deriving the effective Hamiltonian via repeated commutation of the  time independent Hamiltonian with operator $F(t)=-\int_0^t dt' H_{sh}(t)$. Using this approach one can verify that, already in the second commutator, terms containing periodic changes in hopping parameters become negligibly small; that allows us to omit them from our considerations. 

As mentioned in the main text we  invoke instead a time-dependent unitary transformation $\hat{U}=\exp [- {\rm i}\hat H_{\rm int} t - {\rm i} \int^t_0 \hat H_{\rm sh}(t')dt']$. We include in the transformation also the on-site terms with the aim of locating resonant coupling between bands. We obtain the transformed Hamiltonian, $\hat H'=\hat{U}^\dagger\hat  H \hat{U} - {\rm i}\hat{U}^\dagger [d_t \hat{U}]$
in the form: 

\begin{eqnarray}
&&\hspace{-3cm}\hat H' =\left [\exp\left({\rm i}\frac{K}{\omega}\sin \omega t \right)\sum_j \exp\left({\rm i}U_0(n^\ua_j-n^\ua_{j+1})t \right)\left( J_{0}\left [\hat{s}^\dagger_{j+1}\hat{s}_{j}+ \hat{s_\ua}^\dagger_{j+1}\hat{s_\ua}_{j}\right]+ 
T_0 \hat{s_\ua}^\dagger_{j+1}(\hat{n}^{}_ {j+1}+\hat{n}^{}_ {j})\hat{s_\ua}_{j} \right)+ h.c.\right] \nonumber\\
&&\hspace{-2.5cm}+\left [\exp\left({\rm i}\frac{K}{\omega}\sin \omega t \right)\sum_j \exp\left({\rm i}U_1(n^\ua_j-n^\ua_{j+1})t \right)  
\left(J_1\hat{p}^\dagger_{j+1}\hat{p}_{j}+T_1{\hat p_{j+1}}^{\dagger}(\hat{n}^\ua_ {j+1}
+\hat{n}^\ua_ {j})     \hat{p}_{j}   \right)+ h.c.\right] \\
&+& T_{01} \sum_{\langle {ij}\rangle}(j-i){\exp\left[ {\rm i} (E_1+
U_{1}\hat{n}^\ua_i)t + {\rm i} \left ( (i-j)\frac{K}{\omega}+\frac{\delta E_1}{\omega} \right ) \sin \omega t \right] \hat p_{{i}}}^{\dagger}\hat{n}^\ua_ {i} \hat s_{{j}}^{} \nonumber\\
&+& T_{01}  \sum_{\langle {ij}\rangle}(j-i){\exp\left[ {\rm -i} (E_1+
U_{1}\hat{n}^\ua_i)t - {\rm i} \left ( (i-j)\frac{K}{\omega}+\frac{\delta E_1}{\omega} \right ) \sin \omega t \right] \hat s_{{j}}}^{\dagger}\hat{n}^\ua_ {i} \hat p_{{i}}^{}.\nonumber
\end{eqnarray}
Now we assume the resonant condition 
\begin{equation}
E_1+U_1=N\omega + 2\Delta,
\label{eqres}
\end{equation} 
 where $N$ is integer and $\Delta \ll \omega$ is the detuning.
The shaking frequency $\omega$ is chosen  large compared to all the tunneling amplitudes. Before standard time averaging one more simplification is made. We consider low energy Hilbert subspace, where due to strong attractive interactions all $\ua$-fermions are paired. The resonant condition (\ref{eqres}) may be fulfilled  only for sites occupied by the composites, i.e with
$n^\ua_i= \langle \hat n^\ua_i \rangle =1$.  We  average the Hamiltonian  over the oscillation period and neglect terms ($\sim 1/\omega$) obtaining  $H_{\rm eff}$ \eqref{Heff}.
Let us note also that on-site direct excitation of the $p$-band due to periodic shaking (see e.g. \cite{Lacki13}) is negligible in our model due to the resonance condition  (\ref{eqres}) involving composite binding energy.


\section*{References}

\begin{thebibliography}{10}

\bibitem{Su}
A. J. Heeger, S. Kivelson, J. R. Schrieffer, and W. -P. Su, Rev. Mod. Phys. \textbf{60},
781 (1988). 
\bibitem{Zak} J. Zak,  Phys. Rev. Lett. \textbf{62}, 2747 (1989).
\bibitem{Niu}  D. Xiao, M.-C. Chang, and Q. Niu, Rev. Mod. Phys. \textbf{82}, 1959 (2010).
\bibitem{SSH}
 W. P. Su, J. R. Schrieffer, and A. J. Heeger, Phys. Rev. Lett.
\textbf{42}, 1698 (1979).
\bibitem{Jack}
R. Jackiw, and J. R. Schrieffer, Nuclear Phys. B \textbf{190}, 253 (1981).
\bibitem{RM}
M. J. Rice and E. J. Mele, Phys. Rev. Lett. {\bf 49}, 1455 (1982)
\bibitem{RMBloch}
M. Atala, \textit{et. al.}, Nature Physics {\bf 9}, 795(2013)
\bibitem{ssh_superlattice}
F. Grusdt, M. H\''oning, and M. Fleischhauer,  Phys. Rev. Lett. \textbf{110}, 260405 (2013)
\bibitem{Chin}
J. K. Chin et. al., Nature 443, 961 (2006).

\bibitem{Strohmaier}
N. Strohmaier et. al., Phys. Rev. Lett. 99, 220601 (2007).

\bibitem{Hacker}
L. Hackerm\"uller et. al., Science 327, 1621 (2010).

\bibitem{shake0} H. Lignier et al., Phys. Rev. Lett. {\bf 99}, 220403 (2007).
\bibitem{shaking1}
Aidelsburger et al., arXiv:1407.4205 (2014).
\bibitem{shaking2}
Aidelsburger et al., Phys. Rev. Lett. {\bf 111}, 185301 (2013).
\bibitem{shaking3}
Struck et al., Phys. Rev. Lett. {\bf 108}, 225304 (2012).
\bibitem{shaking4}
Struck et al., Nature Physics {\bf  9}, 738-743 (2013).
\bibitem{shaking5}
Miyake et al., Phys. Rev. Lett., {\bf 111}, 185302 (2013).

\bibitem{Andre}
A. Eckardt, C. Weiss, and M.Holthaus, Phys. Rev. Lett. {\bf 95}, 260404 (2005).
\bibitem{triangle}
A. Przysi\k{e}\.zna, O.Dutta, and J. Zakrzewski, 
arXiv:1405.2565 [cond-mat.quant-gas] (2014).

\bibitem{Dut}
O. Dutta, A. Przysi\k{e}\.zna, and M. Lewenstein, Phys. Rev. A \textbf{89}, 043602 (2014).

\bibitem{Dutta11} O. Dutta et al., New J. Phys. \textbf{13}, 023019 (2011).

\bibitem{Mering} A. Mering and M. Fleischhauer,  Phys. Rev. A \textbf{83}, 063630  (2011). 

\bibitem{dirk12} D.-S. L\"uhmann, O. J\"urgensen, and K. Sengstock, New J. Phys.
\textbf{14}, 033021 (2012).

\bibitem{ropp} O. Dutta et al., arXiv:1406.0181 (2014).

\bibitem{SA}
    S. Kirkpatrick,  C. D. Gelatt Jr., M. P. Vecchi, Science {\bf 220}   671-680 (1983)
\bibitem{Lieb72} E.H. 
Lieb and D. Robinson  Commun. Math. Phys. \textbf{28}, 251 (1972).



\bibitem{Shen}
S.-Q. Shen, Topological Insulators: Dirac Equation in Condensed Matters, Springer (Berlin Heidelberg) 2012. 

\bibitem{Ullmo} P. Delplace, D. Ullmo, and G. Montambaux, Phys. Rev. B{\bf 84}, 195452 (2011).

\bibitem{frac}
I. J. R. Aitchison, and G. V. Dunne, Phys.Rev.Lett. \textbf{86} 1690 (2001).
\bibitem{frac1}
G. V. Dunne, and K. Rao, Phys. Rev. D \textbf{64}, 025003 (2001).

\bibitem{AllenEberly} L. Allen and J. H. Eberly,  Optical Resonance and Two-level Atoms, Wiley(New York) 1975.

\bibitem{Mermin} N. W. Ashcroft and N. D. Mermin, Solid State Physics, Holt, Rinehart and Winston 1976.

\bibitem{Jin1}
J. T. Stewart, J. P. Gaebler, and D. S. Jin, Nature \textbf{454}, 744 (2008).

\bibitem{Lew}
M. Lewenstein, A. Sanpera, and V. Ahufinger, Ultracold Atoms in Optical Lattices: Simulating quantum many-body systems,
Oxford University Press, London, (2012). 

 \bibitem{Olshani}
M. Olshanii, Phys. Rev. Lett. \textbf{81}, 938 (1998)

 \bibitem{Floquet}
A. Hemmerich,  Phys. Rev. A {\bf 81}, 063626 (2010)

\bibitem{Lacki13} M. \L{}\k{a}cki and J. Zakrzewski, Phys. Rev. Lett. \textbf{110}, 065301 (2013).



\end{thebibliography}
\end{document}